\documentclass[12pt]{iopart}
\usepackage{hyperref}
\usepackage{amssymb}
\usepackage{graphicx}

\begin{document}

\title{Dynamic Normalization for Compact Binary Coalescence Searches in Non-Stationary Noise}

\author{S. Mozzon$^{1}$, L. K. Nuttall$^{1}$, A. Lundgren$^{1}$, T. Dent$^{2}$, S. Kumar$^{3,4}$, A. H. Nitz$^{3,4}$}
\address{$^{1}$ University of Portsmouth, Portsmouth, PO1 3FX, United Kingdom}
\address{$^{2}$ Instituto Galego de F\'{i}sica de Altas Enerx\'{i}as, Universidade de Santiago de Compostela, 15782 Santiago de Compostela, Galicia, Spain}
\address{$^{3}$ Max-Planck-Institut f{\"u}r Gravitationsphysik (Albert-Einstein-Institut), D-30167 Hannover, Germany}
\address{$^{4}$ Leibniz Universit{\"a}t Hannover, D-30167 Hannover, Germany}

\ead{simone.mozzon@port.ac.uk}

\begin{abstract}
The output of gravitational-wave interferometers, such as LIGO and Virgo, can be highly non-stationary. Broadband detector noise can affect the detector sensitivity on the order of tens of seconds. Gravitational-wave transient searches, such as those for colliding black holes, estimate this noise in order to identify gravitational-wave events. During times of non-stationarity we see a higher rate of false events being reported. 
To accurately separate signal from noise, it is imperative to incorporate the changing detector state into gravitational-wave searches. 
We develop a new statistic which estimates the variation of the interferometric detector noise. 
We use this statistic to re-rank candidate events identified during LIGO-Virgo's second observing run by the PyCBC search pipeline. 
This results in a 5\% improvement in the sensitivity volume for low mass binaries, particularly binary neutron stars mergers. 
\end{abstract}


\maketitle

\section{Introduction}
The Laser Interferometer Gravitational-wave Observatory (LIGO)~\cite{2015CQGra..32g4001L} and Virgo~\cite{TheVirgo:2014hva} have so far observed 14 distinct gravitational-wave signals, two from the merger of binary neutron star systems and twelve from the merger of binary black hole systems~\cite{LIGOScientific:2018mvr, Abbott:2020uma, LIGOScientific:2020stg, Abbott:2020khf}. These three detectors operated at 60-75\% of their design sensitivity in the third observing run (O3)~\cite{Abbott:2020obsscen} which spanned from April 2019 - March 2020. Approximately 50 candidate gravitational-wave events were identified~\cite{LIGO:2020gracedb} in this run. After upgrades these detectors, in addition to KAGRA~\cite{Somiya:2011np,Aso:2013eba}, will return online in 2022 at design sensitivity.

The sensitivity of these interferometers to astrophysical signals is deduced by estimating the power spectral density (PSD) of the noise contributions to the measured strain~\cite{Martynov:2016fzi}. These noise contributions are assumed to be a stationary Gaussian process. In reality however, the output from these instruments 
is highly non-stationary; the detector sensitivity can change on the order of seconds due to broadband sources of noise,  whether instrumental or environmental in origin. 

Searches for coalescing binaries using matched-filtering have contributed to the detection of all transient gravitational-wave signals \cite{LIGOScientific:2018mvr, Nitz_2020, Venumadhav:2019lyq}. Matched-filtering correlates the detector data with a set of possible compact binary coalescence (CBC) waveforms and reweights this correlation with the detector's estimated PSD. This produces a signal-to-noise ratio (SNR) timeseries for each waveform. In the presence of stationary and Gaussian noise the matched filter is the optimal detection strategy~\cite{Allen:2005fk}. Typical searches estimate a detector's PSD over several minutes \cite{Usman:2015kfa, PhysRevD.95.042001, Venumadhav:2019tad}. However, the detector noise can be highly variable over much shorter periods of time. This means the matched filter will not accurately capture the variable nature of interferometric noise, leading to a reduction in search sensitivity. 

In this paper we investigate how incorrectly estimating a detector's PSD affects the detectability of gravitational-wave signals from merging neutron stars and black holes. This is similar to the work presented in Ref~\cite{Zackay:2019kkv}. We develop techniques to account for the changing detector noise and incorporate these methods in to the PyCBC library~\cite{Allen:2005fk, Usman:2015kfa, pycbc-github}. We investigate the impact on the PyCBC search background on data from LIGO-Virgo's second observing run (O2) \cite{Abbott:2019ebz}. We find the sensitivity of PyCBC to binary neutron stars is improved by using techniques to dynamically normalize PyCBC's ranking statistic. However we find no significant improvement to the search's sensitivity for higher mass systems. 

This paper is organized as follows: Section~\ref{sec:background} provides an overview of the matched filter in the presence of non-stationary detector noise. Section~\ref{sec:psdvariation} details the methods used to track non-stationarity in real data. 
In Section~\ref{sec:realdata} we identify and discuss noise variations in LIGO-Hanford during O2.
Section~\ref{sec:snrnormalisation} describes how the PyCBC ranking statistic is dynamically re-normalized to account for PSD variation. Finally, we demonstrate the improvements to the PyCBC search background and therefore sensitivity in O2 as a result of these methods.

\section{Effects of non stationarity in gravitational-wave signal extraction}\label{sec:background}

The mis-estimation of the noise power spectral density due to non-stationarity affects the search for gravitational-wave signals. 
The output of a gravitational-wave interferometer is a time series $s(t)$ such that:
\begin{equation}
    s(t)= \cases
        {n(t)+h(t), & if a signal is present,\\
        n(t), & otherwise,}
\end{equation}
where $n(t)$ is the detector noise and $h(t)$ is a gravitational-wave signal. The noise is assumed to be well described as stationary colored Gaussian noise with zero mean and one-sided PSD, $S_A(|f|)$, defined by
\begin{eqnarray}\label{PSD}
\langle n^{*}(f) n(f') \rangle = \frac{1}{2} S_A (|f|) \delta(f - f') ~,
\end{eqnarray}
where the angle brackets denote averaging over different ensembles of the noise. We use the variables $t$ and $f$ to indicate whether a quantity is in the time or frequency domain. 
Gravitational-wave transient searches estimate the detector noise to construct a SNR time series of the data and identify signals. This is done through the matched filter, which consists of the inner product of the data with the template $h$: 
\begin{eqnarray}
( s | h ) \equiv 2 \int_{-\infty}^{\infty} \frac{s(f) ~ h^{*}(f)}{S_E(|f|)} ~ \mathrm{d}f
\end{eqnarray}
which is appropriate for real-valued data, for which $s^{*}(f) = s(-f)$. Note that the estimated power spectrum of the noise $S_E(|f|)$ is an average of the noise spectrum over a certain time window.
To obtain the SNR time series of the data we normalize the matched filter. If the data do not contain any signals $s(t)=n(t)$, the mean value of the matched filter over different noise realizations cancels, i.e. $\langle ( n | h ) \rangle = 0$ . In this case, the variance of the matched filter is given by
\begin{eqnarray}
\langle |( n | h )|^2 \rangle &=& \Big\langle 4 \int_{-\infty}^{\infty} \int_{-\infty}^{\infty} \frac{n(f) h^{*}(f)}{S_E(|f|)} \frac{h(f') n^{*}(f')}{S_E(|f'|)} ~ \mathrm{d}f \mathrm{d}f' \Big\rangle \\
&=& 2 \int_{-\infty}^{\infty} \frac{|h(f)|^2}{S_A(|f|)} \left( \frac{S_A(|f|)}{S_E(|f|)} \right)^2 ~ \mathrm{d}f ~, \label{var}
\end{eqnarray}
where the second equality follows from the definition of $S_A$, which is the actual spectrum of the noise.
If $S_E\equiv S_A$, (\ref{var}) reduces to $( h | h )$, which motivates the definition of the SNR as
\begin{eqnarray}\label{snr}
\rho \equiv \frac{( s | h )}{\sqrt{( h | h )}} ~.
\end{eqnarray}
It follows that if nothing but noise is present in the data the variance of the SNR is
\begin{eqnarray}\label{Variance}
 \langle\rho^2\rangle =
\left( \int_{-\infty}^{\infty}\frac{|h(f)|^2}{S_E(f)}~ \frac{S_A(f)}{S_E(f)} ~ \mathrm{d}f \right) /
\left( \int_{-\infty}^{\infty}\frac{|h(f)|^2}{S_E(f)} ~ \mathrm{d}f \right) ~.
\end{eqnarray}
Any non-stationarity in the data will lead to a divergence between the estimated spectrum ($S_E$) and the actual one ($S_A$). 
An estimate of the SNR variance can therefore be used to develop a new statistic which measures the variation of the noise PSD. We will refer to this as the \textit{PSD variation statistic}.

\subsection{Optimality of the matched filter}
The mis-estimation of the noise spectrum affects both the inner product of the signal with the template and its variance in noise. Both must be taken into account. For instance, we could use a very narrow bandwidth to define the inner product. This would lower the variance of the inner product in noise, but the matched filter would become less effective overall due to the decrease in the inner product of the signal with the template.
The optimal configuration is obtained when the ratio between the SNR squared and its variance in noise is maximized. 
To prove that this is consistent with $S_E = S_A$ it is convenient to define
\begin{eqnarray}
w(f) = 2 \frac{|h(f)|^2}{S_A(|f|)} \qquad \mathrm{and} \qquad u(f) = \frac{S_A(|f|)}{S_E(|f|)}
\end{eqnarray}
so that
\begin{eqnarray}
( h | h ) &=& \int u(f) w(f) ~df ~, \label{1}\\
\langle ( s | h )^2 \rangle &=& \int u^2(f) w(f) ~df ~. \label{2}
\end{eqnarray}
By considering a signal with unknown amplitude $s = \mathcal{A} h$ and ignoring the noise contribution, the SNR squared is
\begin{eqnarray}
\rho^2 = \frac{(s | h)^2}{(h | h)} = \mathcal{A}^2 (h | h) = \mathcal{A}^2 \int u(f) w(f) ~df ~.
\end{eqnarray}
Using (\ref{1}) and (\ref{2}) we can rewrite the variance of the SNR squared in noise as
\begin{eqnarray}
\langle \rho^2 \rangle = \frac{\langle ( n | h )^2 \rangle}{( h | h )} = \frac{\int u^2(f) w(f) ~df}{\int u(f) w(f) ~df} ~.
\end{eqnarray}
Treating $w(f) df$ as a measure $dw$, the SNR normalized by its variance becomes
\begin{eqnarray}
\frac{\rho_s^2}{\langle \rho^2 \rangle} = \mathcal{A}^2 \frac{\left( \int u \, dw \right)^2}{\int u^2 dw} \leq \mathcal{A}^2 \int dw ~,
\end{eqnarray}
where the Cauchy-Schwarz inequality is used to set an upper limit to the ratio. The limit is saturated when $u = 1$, i.e. when the estimated spectrum is equal to the actual one. Accordingly, if data are non-stationary the inner product is not an optimal receiver and therefore the matched filter SNR does not provide the optimal detection statistic.

\section{PSD variation of gravitational-wave data}\label{sec:psdvariation}
We estimate the variation of the noise PSD with a similar approach as Ref.~\cite{Zackay:2019kkv}. We model the relation between the actual spectrum and the estimated one with a frequency-independent factor $v_s$, such that $S_A = v_s S_E$. We define the PSD variation as the time series which tracks $v_s$. 

To compute $v_s(t)$ we estimate the SNR variance of the data as a function of time.  We begin by splitting the data into 512 seconds segments. For each segment we compute $S_E(f)$ using the Welch method on overlapping segments of 8 seconds. This is identical to how the PyCBC search estimates noise.  We then create a filter 
\begin{eqnarray}\label{filter}
\mathcal{F}(f) = \frac{|h(f)|}{S_E(f)}
\end{eqnarray} to be applied to the data. Here we use an approximate expression for CBC templates which captures the dominant amplitude behavior, giving $|h(f)| \propto f^{-7/6}$~\cite{PhysRevD.80.084043}.  To remove undesired frequencies from the data we combine $\mathcal{F}(f)$ with a bandpass filter. The low frequency threshold (20 Hz) is determined by the LIGO detector sensitivity, while frequencies above 480 Hz are filtered out to remove strong spectral lines~\cite{2015CQGra..32g4001L} and to reduce the computational cost. The combination of filters is then smoothed and convolved with the data. The spectrum of the resulting time series, integrated over frequency, gives an estimation of $\langle\rho^2\rangle$ up to a constant. Using Parseval's theorem, we can then compute the PSD variation, $v_s$ at a given time $t_0$ as
\begin{eqnarray}\label{PSD_variation}
v_s(t_0)\equiv \mathcal{N} ~ \langle\rho^2\rangle  (t_0) \simeq \mathcal{N} \int_{t_0-\Delta t}^{t_0}|\mathcal{F}(t)*s(t)|^2 \mathrm{d}t ~,
\label{eqn:vs}
\end{eqnarray}
where $\mathcal{F}(t)*s(t)$ is the convolution between the filter and the data and $\mathcal{N}$ is a constant such that the expectation value of $v_s$ in Gaussian stationary noise is 1. The resulting time series corresponds to the estimated SNR variance of the previous $\Delta t$ seconds. Therefore, $\Delta t$ determines the typical time scale at which we want to track variations in the noise PSD. Because data is a discretely sampled quantity, instead of integrating we compute $v_s$ through the mean square of the filtered data.

Our estimate can be strongly affected by loud instrumental noise transients, known as glitches \cite{TheLIGOScientific:2016zmo,2018RSPTA.37670286N}, which corrupt the data over short time scales. To reduce this contribution we perform a preliminary data cleaning step: we first compute (\ref{eqn:vs}) over a short timescale of $\Delta t$ = 0.25s then identify any outliers in the resulting timeseries $v_s(t_i)$ that are greater than twice the mean of the surrounding values, $0.5(v_s(t_{i-1}) + v_s(t_{i+1}))$. Outlier values are substituted with the corresponding mean value, yielding a cleaned time series that will then be averaged over a longer timescale to produce the final estimate of $v_s(t)$.

Using a set of simulated merger signals we investigate how the presence of real gravitational-wave events in the data affect the PSD variation statistic.
We simulated a population of compact binary mergers \cite{STT4, Bohe:2016gbl, Husa:2015iqa} uniform in chirp mass and distance which spans the astrophysical signal space observable by LIGO. We added these signals to 4 hours of simulated stationary Gaussian noise and computed the PSD variation. For signals with SNR below 15, the PSD variation distribution does not change due to the presence of the signals; none of the PSD variation values for simulated mergers are distinguishable from the values we obtained for noise. For louder sources the PSD variation scales linearly with the square of the signal SNR. In particular, a detection statistic obtained by re-scaling $\rho$ with $\sqrt{v_s(t)}$ is a monotonically increasing function of the SNR for any CBC signal. Therefore, the ranking statistic of loud sources, even if rescaled, remains above the detection threshold.

\subsection{Uncertainty in the estimation}
Tracking variations of the PSD over short time scales increases the uncertainty of the estimation of the SNR variance. If the data are stationary Gaussian noise we can gather the statistical properties of our estimator analytically. In this case, it is convenient to factor the filter $\mathcal{F}(f)$ into a  weighting and a whitening factor. The filter applied to the data in the frequency domain is therefore
\begin{eqnarray}
\mathcal{F}(f)s(f) = \frac{|h(f)|}{\sqrt{S_E(f)}} ~ \frac{s(f)}{\sqrt{S_E(f)}} ~, 
\end{eqnarray}
where the final factor is approximately white Gaussian noise. This factor has unit-variance Gaussian real and imaginary parts in each frequency bin. Our statistic is a sum over frequencies of $|\mathcal{F}(f)s(f)|^2$ . The squared magnitude of the final term is a $\chi^2$ distribution with two degrees of freedom, so our statistic is a weighted sum of $\chi^2(2)$ random variables.
The moment generating function of the distribution is $M_t = \prod_{j=1}^N(1-2w_jt)^{-1}$, where $w_j = \frac{|h(f_j)|^2}{S_E(f_j)}$ is the weight of the j-th frequency bin and $N$ is the total number of frequency bins. Accordingly, the variance of the PSD variation distribution relative to its mean is
\begin{eqnarray}\label{variance}
        \frac{\mathrm{Var}(v_s)}{\langle v_s \rangle^2}=\frac{\sum_{j=1}^Nw_j^2}{\left(\sum_{j=1}^Nw_j\right)^2} \equiv \frac{1}{N_{\rm eff}} = \frac{1}{\Delta t ~ \Delta f_{\rm eff}} \ge \frac{1}{N}
\end{eqnarray}
where the lower limit is defined by the Cauchy-Schwarz inequality and is reached when the weights $w_j$ are all equal. $\Delta f_{\rm eff}$ is the number of weights which contributes to the summation per second and it represents the frequency range where the detector is effectively sensitive to gravitational-wave signals. 
For the current LIGO detectors we expect the dominant weights to be in a frequency range of approximately 100 Hz, due to the detector sensitivity. 
To estimate the variance we fix the integration time, $\Delta t$, to 8 seconds. This time window is consistent with the typical time-scale of the noise variations and it is sufficient to encompass the main power of typical CBC waveform templates. 
Using (\ref{variance}) we then obtain a variance of the PSD variation distribution relative to its mean of $1.25 \times 10^{-3}$. To verify our predictions, we computed the PSD variation over 4 hours of simulated LIGO-Hanford detector noise. We measure a value of $1.7 \times 10^{-3}$. The effective bandwidth of the Hanford detector is therefore approximately 70 Hz. We find similar bandwidths for LIGO-Livingston and Virgo.

Accounting for noise variations will be increasingly important in the future, when the ground based gravitational-wave detectors reach their design sensitivity. Moreover, the next generation of ground based detectors, such as the Einstein Telescope \cite{Punturo:2010zza}, will have a higher bandwidth which allows tracking of non-stationarity over shorter time scales.

\section{PSD Variation Statistic over example O2 Data}\label{sec:realdata}

\begin{figure}\label{real_data}
\includegraphics[width=0.5\textwidth]{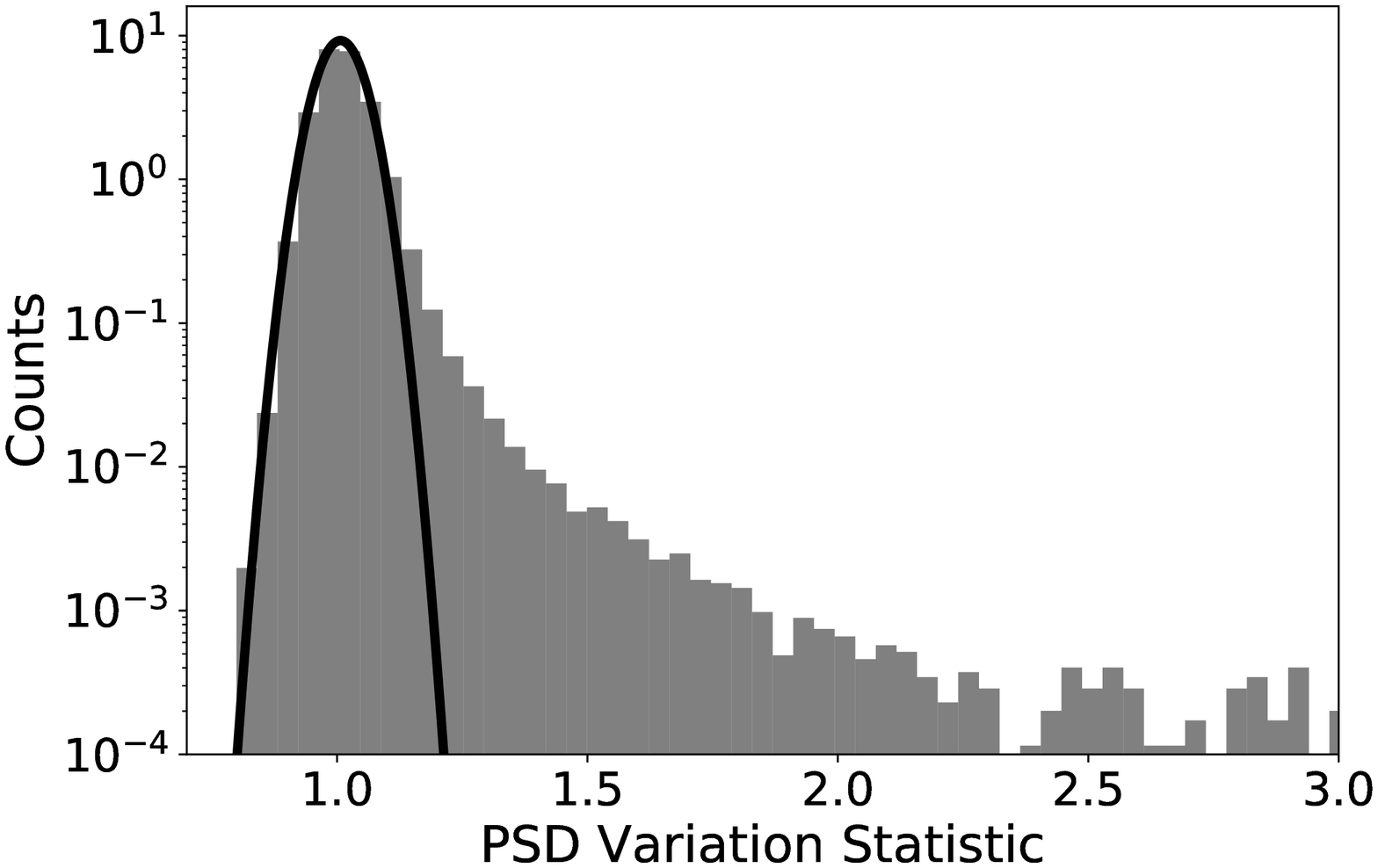}\includegraphics[width=0.5\textwidth]{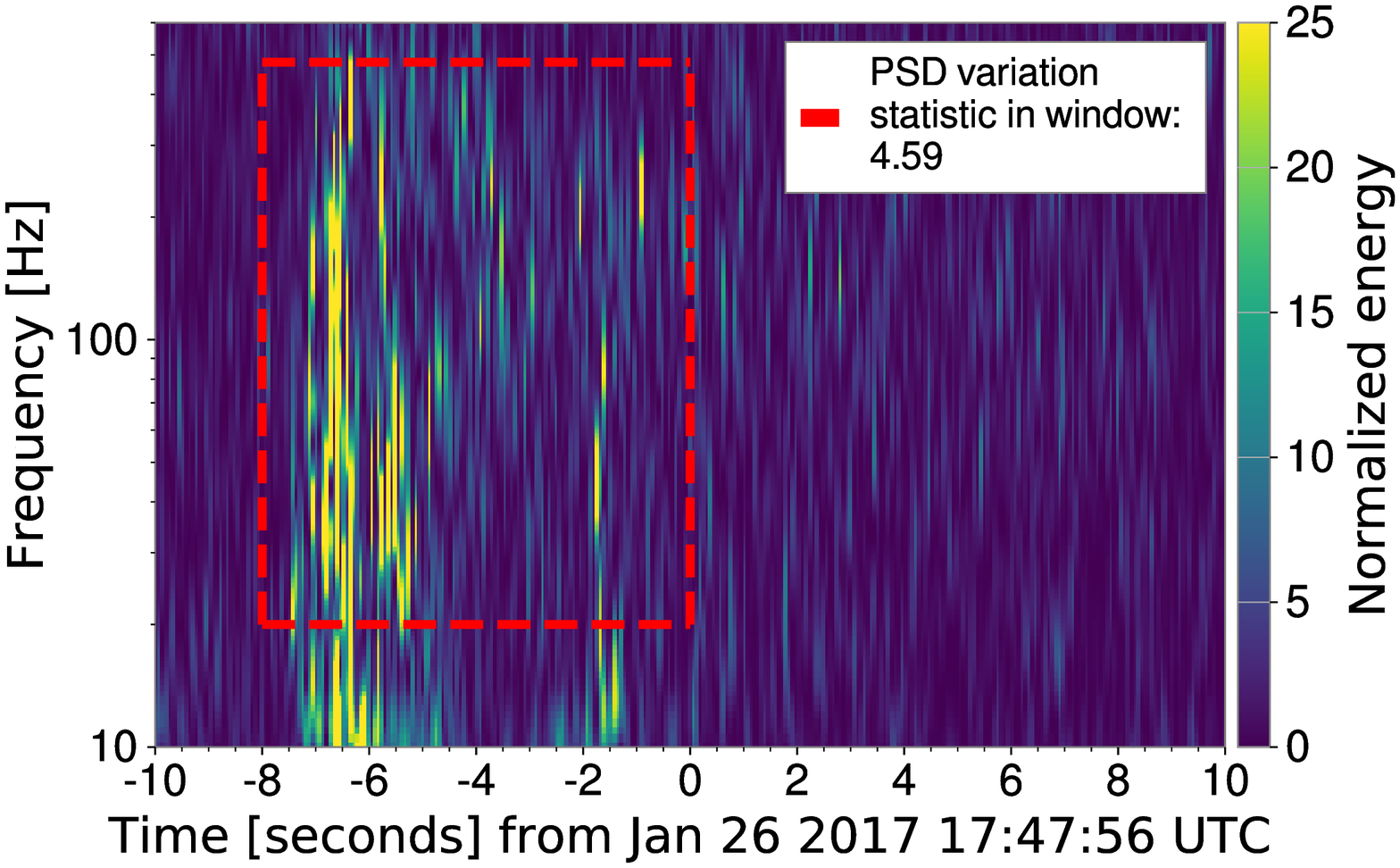}
\caption{\label{Fig:real_times} (\textit{Left}) Distribution of the PSD variation statistic of LIGO-Hanford data between January 22th 2017 08:00:00 UTC and February 3rd 2017 16:20:00 UTC. The black line shows the expected distribution for Gaussian stationary noise. We restrict the x-axis to 3 for visualization purposes. 
(\textit{Right}) Time-frequency spectrogram of data belonging to the tail of the PSD variation distribution. The red dashed box encloses the data used to compute the PSD variation statistic at the reference time. The associated PSD variation value is reported in the label.}
\end{figure}

We use the SNR variance to track times of non-stationarity of LIGO data collected during the second observing run. We consider LIGO-Hanford data between January 22th 2017 08:00:00 UTC and February 3rd 2017 16:20:00 UTC, corresponding to $\sim 10$ days of observing data. During this period the noise spectrum of the detector was extremely unstable due to environmental disturbances, such as high micro-seismic noise due to bad weather conditions. Consequently, we see wide variations in the detector sensitivity.

Figure \ref{Fig:real_times} shows the distribution of the PSD variation $v_s(t)$ during this period compared to the theoretical prediction from Gaussian noise. The Hanford data clearly shows non-stationarity which appears as a tail of high PSD variation values. In Figure \ref{Fig:real_times} we have restricted the x-axis for visualization purposes. In fact, the tail of the distribution extends to PSD variation values of order of $10^3$. These instances are due to a class of extended clusters of transient noise which are relatively rare and are not the target of this method. Data which are not consistent with Gaussian noise represent $\sim3\%$ of the whole period. 

The right plot of Figure \ref{Fig:real_times} shows a time-frequency representation of example data picked from the tail of the distribution. The power excesses, shown in yellow, are clearly not associated to any gravitational-wave signal and their origin is not trivial. Within the tail are other forms of noise which can look very visually different to Figure \ref{Fig:real_times}. Our method has the advantage of identifying many different forms of noise that can impact a matched-filter search. We therefore developed a PSD variation monitor to assess the stationarity of the LIGO detector data in low latency to inform possible retractions of false LIGO-Virgo events. This low latency monitor has been running for the entirety of O3.

\section{SNR normalisation}\label{sec:snrnormalisation}

Gravitational-wave searches for coalescing binaries based on matched filtering identify signals by correlating the detector data against a set of CBC template waveforms. The searches begin by dividing the data into blocks of several minutes and compute the average noise spectrum of the detectors for each period. These are used to create a matched-filter SNR time series for each template for each detector. Every local maximum in the SNR time series which exceed a fixed threshold defines a \textit{single detector trigger}. These triggers can be generated either by gravitational waves or noise artefacts.
The PyCBC search pipeline down-ranks loud noise transients with a chi-squared test which checks that the accumulation of signal power as a function of frequency is consistent with the matching template waveform. The re-weighted SNR, $\hat{\rho}$, a function of $\rho$ and $\chi^2$ given by (1) of Ref.~\cite{abadie2012search}, defines the detection statistic for a single detector that ranks the likelihood for a trigger to be due to a real signal versus noise. 

Short term fluctuations of the noise spectrum affect the optimality of the match filter and can bias the distribution of triggers, even in otherwise Gaussian noise. However, we can not account for these variations directly during the match filtering because estimating the PSD with sufficient accuracy to maintain search sensitivity requires durations of data of order a few hundred seconds \cite{Allen:2005fk}. 
Moreover, in order to construct the matched filter the same PSD must be used over the duration of the signal. Calculating the PSD over short time periods (i.e. order of seconds) is therefore problematic for long signals like binary neutron stars mergers (i.e. on the order of minutes). A better approach is to re-rank the SNR time series with the PSD variation statistic during times of non-stationarity.

We implemented this approach in the PyCBC search to account for noise variations of O(10s) timescales. In Ref.~\cite{Nitz_2020} a correlation was demonstrated between the PSD variation and the rate of noise triggers above a given threshold in $\hat\rho$. 
The rate was found to be a function of $\hat\rho v_s(t)^{-\kappa}$, where $\kappa$ is a constant allowing for deviation from the expected behavior $\hat\rho v_s^{-1/2}$.  Estimates of $\kappa \gtrsim 0.33$ were obtained, thus a re-scaled statistic $\hat\rho v_s(t)^{-0.33}$ was employed to obtain search results in Ref.~\cite{Nitz_2020}.
The non-ideal estimated value of $\kappa$ likely resulted from a combination of thresholding and clustering effects on the trigger distribution, and from not accounting for PSD variation in the chi-squared test calculation. In this work we consider the effects of non-stationarity in both the SNR time series and the chi-squared test. Since the SNR scales as $S_E^{-1/2}$, we correct the trigger detection statistic dynamically re-scaling the SNR by a multiplicative factor $v_s(t)^{-0.5}$. The chi-squared time series depends on the power of the signal, we therefore re-scale it by $v_s(t)^{-1}$. Accordingly, the re-weighted SNR becomes:
\begin{equation}
    \hat{\rho}_{corr}= \cases
        {\hat{\rho} v_s^{-1/2}, & for $\chi_r^2 v_s^{-1}\leq 1$,\\
        \hat{\rho} \left[ \frac{1}{2}(v_s^3+(\chi_r^2)^3)\right]^{-1/6}, & for $\chi_r^2 v_s^{-1} > 1.$}
\end{equation}

\begin{figure}
\includegraphics[width=0.5\textwidth]{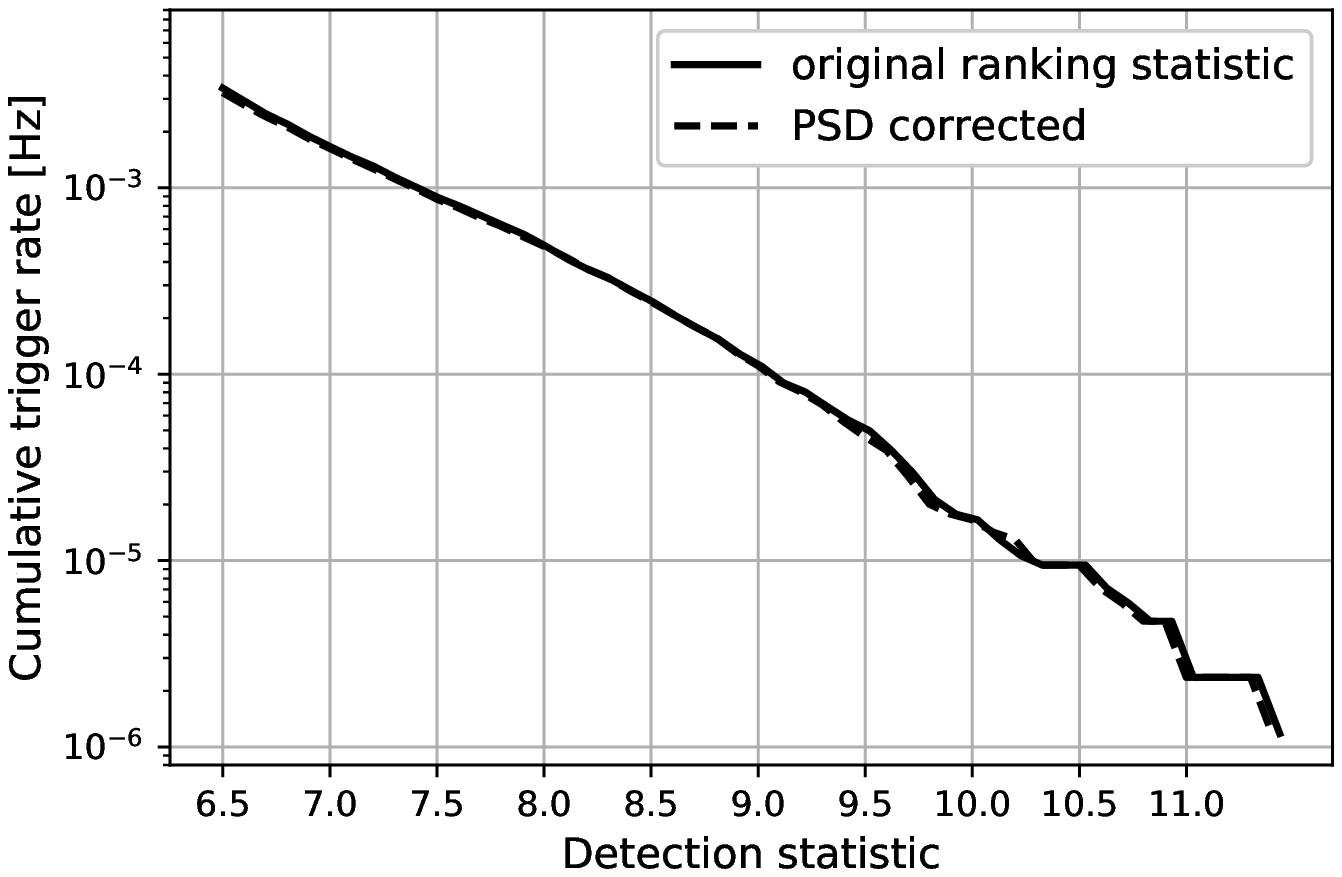}\includegraphics[width=0.5\textwidth]{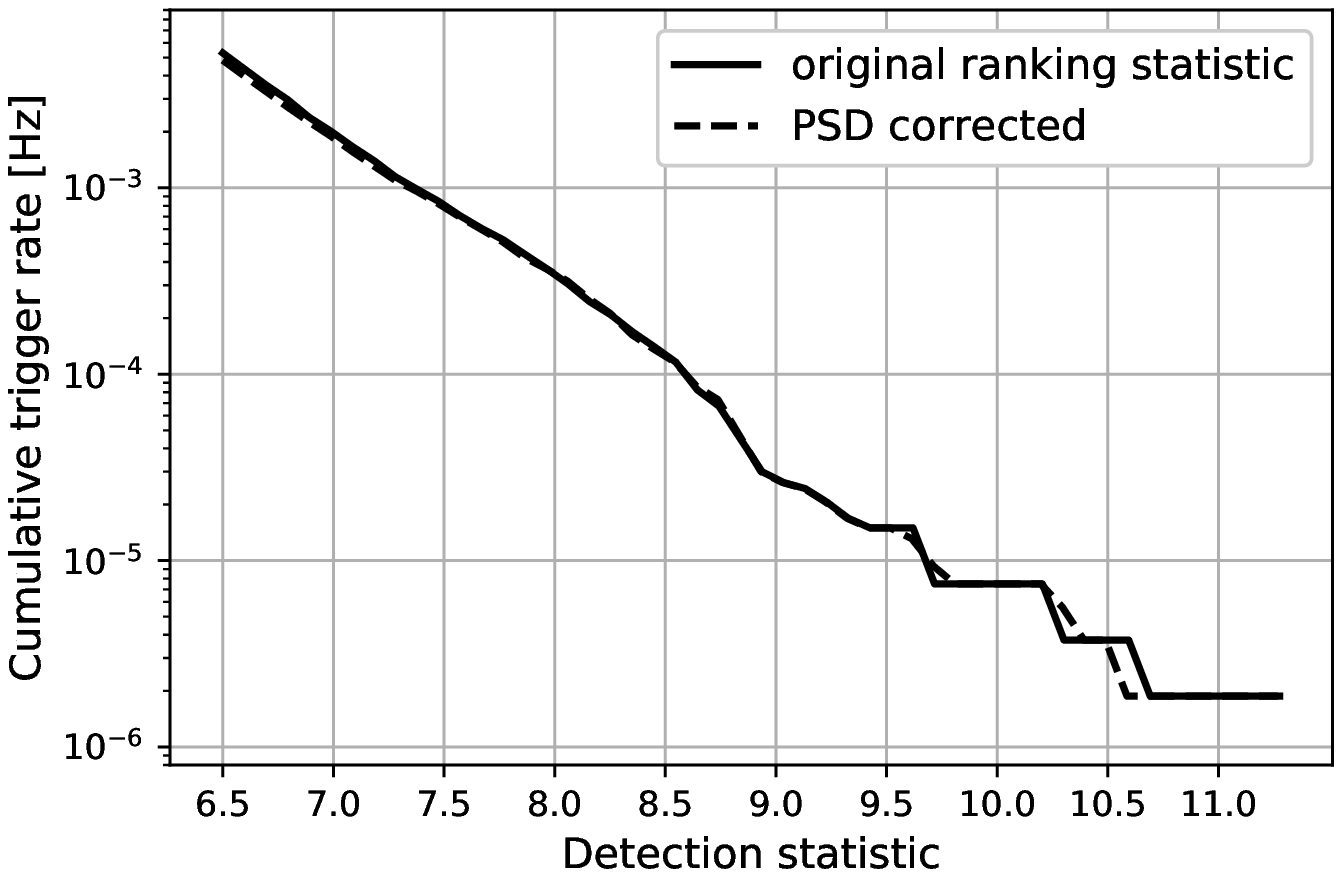}
\includegraphics[width=0.5\textwidth]{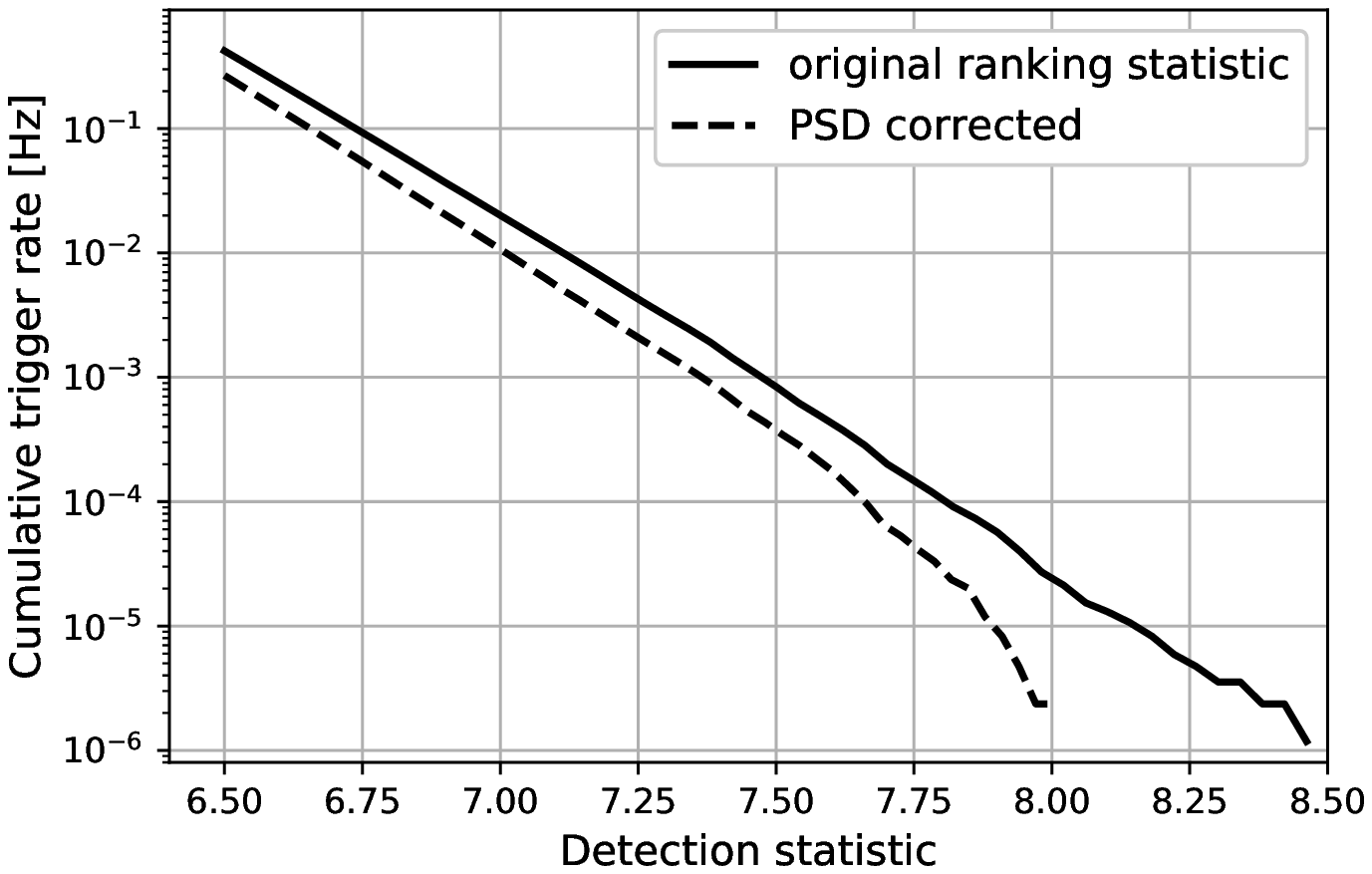}\includegraphics[width=0.5\textwidth]{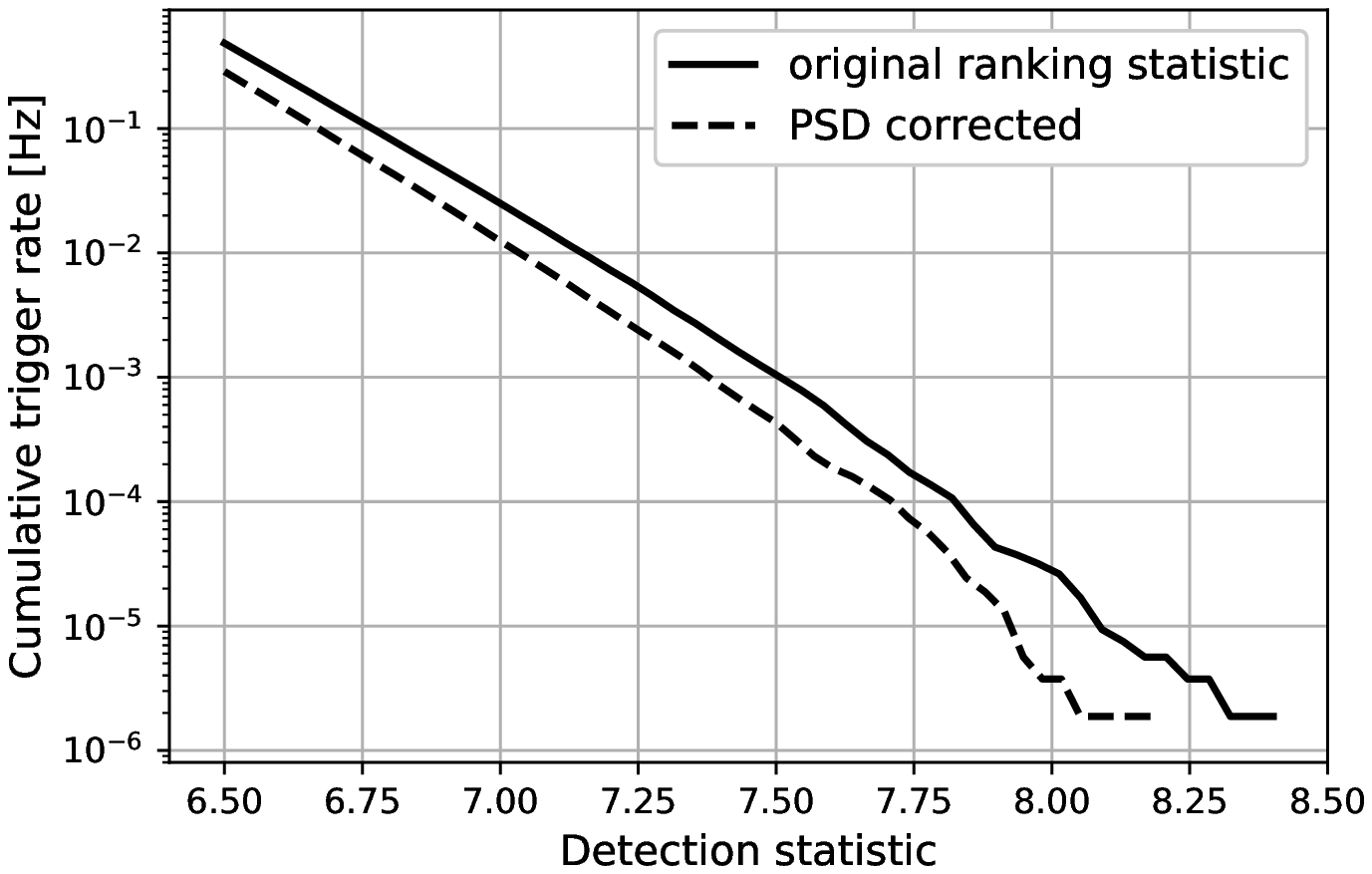}
\caption{\label{Fig:bkg} Cumulative trigger rate computed using LIGO-Hanford data (left plots) and LIGO-Livingston data (right plots) between January 22th 2017 08:00:00 UTC and February 3rd 2017 16:20:00 UTC. Upper plots consider just triggers recovered by short duration templates (0.175 - 0.51 seconds). Triggers in the lower plots use long duration templates (longer than 4 seconds). The single detector background distribution of long templates triggers is strongly affected by the PSD variation correction.}
\end{figure}

In Figure \ref{Fig:bkg} we present how this correction affects the single detector background distribution during the same period considered in Section \ref{sec:realdata}.
Each plot shows the cumulative trigger rate as a function of the detection statistic with and without the PSD variation correction.
We remove triggers generated by real signals to only examine the noise distribution. We compute the duration of the templates using a low-frequency cutoff of 20 Hz.
The distribution of triggers associated to templates with duration longer than 4 seconds changes drastically for both Hanford and Livingston detectors. Short templates (0.175 - 0.51 seconds) however are only slightly affected by our correction. Indeed, the PSD variation statistics is insensitive to isolated short noise transients (glitches), the effect of which averages out over the 8 seconds integration time. Reducing the integration time could improve our ability to downrank short templates, however it increases the uncertainty of the estimation. 
Short duration templates also appear with a very high detection statistic. In the case of Gaussian noise, we would expect to obtain similar distributions between short and long templates. This asymmetry suggests the presence of short noise transients which are not not fully suppressed by the pipeline affecting the sensitivity of the search \cite{Cabero:2019orq}. The background distribution for short template will require further investigations.

To quantify the improvement in the search sensitivity due to the PSD variation correction, we estimate the sensitivity volume of the search. Considering just the two LIGO detectors, we can naively estimate the network coincident trigger rate, $R_N$ as:
\begin{eqnarray}
    R_N = R_{H1}\times R_{L1} \times 2\Delta t \,
\end{eqnarray}
where $R_{H1}$ and $R_{L1}$ are the trigger rate for Hanford and Livingston detectors and $\Delta t$ is the time-distance between them. Noticing that the performance are similar for the two detectors, we can then approximate $R_{H1} \simeq R_{L1}$. Fixing the network false alarm rate (FAR) to 1 per year we then obtain $R_{H1}=R_{L1}=1.25 \times 10^{-3}$ Hz, corresponding to a single detector FAR of 1 per 13 minutes. We finally estimate the volume increase as:
\begin{eqnarray}
    \frac{V_{corr}}{V} = \left[\frac{\hat{\rho}_{corr}|_{FAR=1/13\, mins}}{\hat{\rho}|_{FAR=1/13\,mins}}\right]^{-3}
\end{eqnarray}
where $\hat{\rho}$ is the initial detection statistics for a single detector. We obtain a volume improvement of $\sim 1\%$ for short templates, while for long templates the sensitivity increases by $\sim 5\%$. 

Using the same data, we also compare the search sensitivity of the two detection statistics by adding 20,000 simulated signals during the PyCBC analysis~\cite{Usman:2015kfa}. We simulated a population of mergers isotropically distributed in sky location, spanning uniformly the log-component-mass search space. The fraction of injections identified by the pipeline at a given FAR determines the sensitivity volume of the search. 
This method shows volume improvements due to the PSD variation correction which are compatible with our previous estimates within uncertainties.

\section{Conclusion}
This paper presents a new approach to account for noise variations in interferometric gravitational-wave detectors which can limit the detection of compact binary coalescences. 
We developed a new statistic to track noise variations which is based on estimating the variance of the SNR. We have incorporated this method in the PyCBC search pipeline. In particular, we used our statistic to dynamically re-rank the trigger detection statistic. Analyzing 12 days of O2 LIGO data, we found that our correction significantly changes the single detector background distribution for long waveform triggers. Consequently, we measure a 5\% increase in the sensitive volume of the PyCBC search for binary neutron stars, neutron star-black hole and low mass black hole systems. In contrast, the sensitivity to higher mass systems did not show any significant improvement.

During O3 both LIGO detectors have been strongly affected by non stationary noise, such as scattered light \cite{alog-wa, alog-la}. For this reason we expect our approach to provide much larger sensitivity improvements in the analysis of O3 data.
Accounting for noise variations will be even more essential in the future, when the gravitational-wave detectors will reach their design sensitivity. Moreover, the next generation of ground based detectors will have a higher bandwidth which allows to track non-stationarity over shorter time scales.

\ack
We thank Ian Harry, Connor McIsaac and Derek Davis for useful discussions and suggestions.
This research has made use of data, software and/or web tools obtained from the Gravitational Wave Open Science Center (https://www.gw-openscience.org), a service of LIGO Laboratory, the LIGO Scientific Collaboration and the Virgo Collaboration. LIGO is funded by the U.S. National Science Foundation. Virgo is funded by the French Centre National de Recherche Scientifique (CNRS), the Italian Istituto Nazionale della Fisica Nucleare (INFN) and the Dutch Nikhef, with contributions by Polish and Hungarian institutes. The  authors  are  grateful for  computational resources  provided  by  the  LIGO  Laboratory  and  supported  by  National  Science  Foundation  Grants  PHY-0757058 and PHY-0823459. SM was supported by a STFC studentship. This document has been assigned LIGO Laboratory document number LIGO-P2000069.

\section*{References}
\bibliographystyle{unsrt}
\bibliography{dynamic_normalization}

\end{document}